\documentclass[prb,aps,twocolumn,superscriptaddress,showpacs]{revtex4}
\usepackage{graphicx}
\begin{document}

\title{Neutron scattering search for static magnetism in oxygen
ordered YBa$_{2}$Cu$_{3}$O$_{6.5}$}

\author{C. Stock}
\affiliation{Physics Department, University of Toronto, Toronto,
Ontario M5S 1A7, Canada }
\author{ W.J.L. Buyers}
\affiliation{National Research Council, Chalk River, Ontario, K0J
1J0, Canada}
\affiliation{Canadian Institute of Advanced Research,
Toronto, Ontario, M5G 1Z8, Canada}
\author{ Z. Tun}
\affiliation{National Research Council, Chalk River, Ontario, K0J
1J0, Canada}
\author{R. Liang}
\affiliation{Physics Department, University of British Columbia,
Vancouver, B.C., V6T 2E7, Canada}
\affiliation{Canadian Institute of
Advanced Research, Toronto, Ontario, M5G 1Z8, Canada}
\author{D. Peets}
\affiliation{Physics Department, University of British Columbia,
Vancouver, B.C., V6T 2E7, Canada}
\author{D. Bonn}
\author{W.N. Hardy}
\affiliation{Physics Department, University of British Columbia,
Vancouver, B.C., V6T 2E7, Canada}
\affiliation{Canadian Institute of
Advanced Research, Toronto, Ontario, M5G 1Z8, Canada}
\author{L. Taillefer}
\affiliation{Physics Department, University of Toronto, Toronto,
Ontario M5S 1A7, Canada } \affiliation{Canadian Institute of Advanced
Research, Toronto, Ontario, M5G 1Z8, Canada}

\date{\today}

\begin{abstract}

We present elastic and inelastic neutron scattering results on highly
oxygen ordered YBa$_{2}$Cu$_{3}$O$_{6.5}$ ortho-II. We find no evidence
for the presence of ordered magnetic moments to a sensitivity of $\sim$
0.003 $\mu_{B}$, an order of magnitude smaller than has been suggested
in theories of orbital or d-density-wave (DDW) currents.  The absence
of sharp elastic peaks, shows that the d-density-wave phase is not
present, at least for the superconductor with the doping of 6.5 and the
ordered ortho-II structure.  We cannot exclude the possibility that a
broad peak may exist with extremely short-range DDW correlations.  For
less ordered or more doped crystals it is possible that disorder may
lead to static magnetism. We have also searched for the large normal
state spin gap that is predicted to exist in an ordered DDW phase.
Instead of a gap we find that the Q-correlated spin susceptibility
persists to the lowest energies studied, $\sim6$ meV. Our results are
compatible with the coexistence of superconductivity with orbital
currents, but only if they are dynamic, and exclude a sharp phase
transition to an ordered d-density-wave phase.

\end{abstract}
\pacs{74.72.-h,75.25.+z,75.40.Gb}

\maketitle
\section{Introduction}

The search for the origins of high-temperature superconductivity has
led to many new concepts in condensed matter
science.~\cite{BatloggPhysWld:13p33y00} One of these is that
antiferromagnetism can originate from orbital currents. This idea
underlies the staggered flux phase of Marston and
Affleck~\cite{Marston89:43}, the proposed dynamic orbital currents of
Wen and Lee~\cite{Wen96:76}, and the d-density wave (DDW) order of
Chakravarty {\it{et al}}.~\cite{Chak01:63,Chak01:109,Chak01:204}  The
orbital currents would flow in the planes of cuprate superconductors
around a region of the size of the unit cell.  They would be equivalent
to a small magnetic moment to which neutrons are sensitive. The theory
of Chakravarty {\it{et al.}} predicts that static d-density wave (DDW)
order would appear as an elastic Bragg peak well above the
superconducting transition temperature. Most models require a breaking
of the translational symmetry of the CuO$_{2}$ planes at \textbf{Q} =
($\pi$, $\pi$), but the orbital currents predicted by
Varma~\cite{Varma97:55} do not.

There have been several reports of a static magnetic signal at ($\pi$,
$\pi$) in the YBa$_{2}$Cu$_{3}$O$_{6+x}$ (YBCO) at a surprisingly large
ordering temperature of $\sim$ 300 K.  Sidis \textit{et
al.}~\cite{Sidis01:86} have performed both unpolarized and polarized
neutron scattering measurements along with zero-field $\mu$SR
experiments on YBCO$_{6.5}$. They observed an elastic magnetic peak at
\textbf{Q} = (1/2 1/2 L) for L=2 but not for L=0. This they found to be
consistent with Cu$^{2+}$ moments of $\sim$ 0.05 $\mu_{B}$ pointing in
the \textit{a-b} plane and antiferromagnetically coupled within the
plane and between the bilayers. However, the $\mu$SR results conducted
on a piece from the same sample did not display oscillations consistent
with the magnetic moment observed by neutron scattering.

At a slightly larger oxygen doping Mook \textit{et
al.}~\cite{Mook01:64} observed an elastic peak at (1/2 1/2 2) in
YBCO$_{6.6}$ below 300 K. They derived an effective ordered magnetic
moment of $\sim$ 0.02 $\mu_{B}$ and noted that the form factor drops
rapidly with $|$\textbf{Q}$|$ as would be expected from orbital
currents.  The $\mu$SR results by Sonier \textit{et
al.}~\cite{Sonier01:292} on YBCO$_{6.67}$ and YBCO$_{6.95}$ were
initially interpreted as showing the existence of static magnetism
consistent with the d-density wave or other orbital current theories.
However, further work~\cite{Sonier01:479} has shown that the complex
$\mu$SR results are more likely to arise from the charge
inhomogeneities of a stripe structure rather than from d-density wave
order.  For YBCO$_{6.5}$ in a magnetic field, Miller \textit{et
al.}~\cite{Miller01:550} have described their $\mu$SR results with a
model that gives an improved fit when a static antiferrromagnetic
moment is included in the vortex cores. At doping $x>0.5$ we note that
filled Cu-O chain segments may lie adjacent, while for YBCO$_{6.5}$,
ortho-II ordered, the chains lie two cells apart and an empty Cu chain
separates them.

Chakravarty \textit{et al.}~\cite{Chak01:109} state that the results of
Ref.\ \onlinecite{Mook01:64}.   are consistent with the DDW theory.
This claim is based on three observations.  An elastic magnetic peak
was observed at ($\pi$, $\pi$), its intensity decreases much faster
with $|$\textbf{Q}$|$ than does the Cu$^{2+}$ spin form
factor,~\cite{Shamoto93:48,Tewari01:027} and finally, since the DDW is
Ising-like and breaks a discrete symmetry, there should be a gap, which
they point out is consistent with the $\sim$ 20 meV gap reported by Dai
\textit{et al.} in YBCO$_{6.6}$.~\cite{Dai01:63,Dai98:241}

It is clearly important to know whether static magnetism, regardless of
its microscopic origin, plays a universal role in the superconductivity
of YBCO, and whether it depends on oxygen concentration, oxygen order,
and orthorhombic twinning.  To answer this we have carried out elastic
and inelastic neutron scattering on detwinned orthorhombic YBCO$_{6.5}$
in which the oxygen is well ordered in the ortho-II phase.  In this
phase copper chains along \textit{b*} filled with oxygen occur every
second cell along the \textit{a*} direction. We establish the absence
of elastic magnetic signal with a high sensitivity that would have
easily detected the peaks seen in Ref.\ \onlinecite{Sidis01:86} and
Ref.\ \onlinecite{Mook01:64}.  We also observe low-energy spin
fluctuations in the normal state well below 20 meV where a spin-gap has
been claimed.~\cite{Dai98:241}

\begin{figure}[t]
\includegraphics[width=80mm]{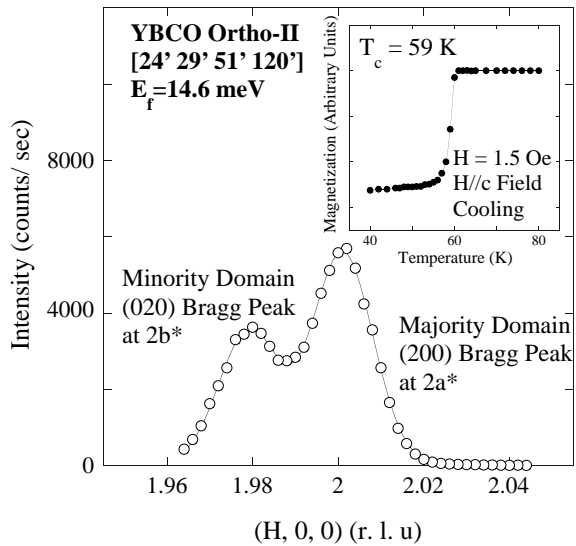}
\caption{Radial scan through the (2 0 0) Bragg peak.  The gaussian fit
shows that the majority domain occupies $70\%$ of the sample volume.
The magnetisation inset shows a sharp superconducting transition
temperature at 59 K.}
\end{figure}

\section{Experiment}
The sample consisted of six orthorhombic~\cite{Jorg90:41,Casalta96:258}
crystals of YBa$_{2}$Cu$_{3}$O$_{6.5}$ grown at the University of
British Columbia using a top-seeded melt growth technique
~\cite{Cima92:179,Bateman92:1281,Izumi93:757} which in common practice
requires the addition of small amounts of platinum and
Y$_{2}$BaCuO$_{5}$ (green phase) to modify growth dynamics.  In our
case 0.5 wt$\%$ of platinum and 2 wt$\%$ of Y$_{2}$BaCuO$_{5}$ were
added, which are at the low range of values used in this technique.  It
was found by EDX composition mapping that platinum exists in the
Ba$_{3}$Y$_{2}$PtCu$_{2}$O$_{10}$ phase and the YBCO matrix is
essentially free of platinum.  By scaling the diffraction intensities
of impurity phases to the weak extinction free YBCO (1 1 2) Bragg peak,
we found that the dominant impurity was the green phase of $\sim5\%$ by
volume; all other impurities were less than $1\%$ by volume.  We note
that green phase fractions as large as 14 $mol\%$ have been reported in
other studies.~\cite{Dai96:77}

    The oxygen content of the crystals was set to 6.5 by annealing at
$760^{o}$C in oxygen flow followed by quenching to room temperature in
nitrogen gas flow.  Partially detwinned crystals were obtained by
mechanically applying a pressure of 100 MPa along the
\textit{a}-direction at $400^{o}$C in nitrogen gas flow.  The ortho-II
ordering, which has alternating full and empty chains, was developed by
annealing the crystals at $60^{o}$C for 2 weeks in a sealed bottle. The
crystals show a sharp superconducting transition at 59 K with a width
of 2.5 K, as observed by field cooling magnetization shown in Fig. 1.
Each crystal, about 1 cm$^{3}$ in size, was sealed in an aluminum can
under a dry helium atmosphere (kept at dewpoint $<$ -40$^{o}$C to
eliminate water). The Stycast sealant was masked with gadolinium paint.
The six crystals were mutually aligned on a multi-crystal mount at the
E3 spectrometer at NRU reactor, Chalk River. The rocking curve width
was about 1$^{o}$ for each crystal and approximately 2.2$^{o}$ for the
composite.

From the (2 0 0) radial scan of Figure 1 we find from a gaussian fit
that the majority domain occupies 70$\%$ of the sample volume. The peak
at higher $|$\textbf{Q}$|$ (H=2) is the (2 0 0) Bragg peak from the
majority domain, and the peak H=1.98 is the (0 2 0) of the minority
domain. An independent check is obtained from the satellites produced
by oxygen chain order peaks, (3/2 0 0) and (0 3/2 0)(Fig. 2). Their
widths along H and K were nearly resolution limited.  Fits to
resolution-convolved lorentzians showed that the oxygen correlation
lengths exceeded $\sim$100 \AA~in both the \textit{a} and \textit{b}
directions, while it was approximately 50 \AA~along the \textit{c}
direction.  These correlation lengths compare favorably to the x-ray
characterization of highly ordered ortho-II indicating the high quality
of our crystals.~\cite{Andersen99:317,Liang00:336} We find that the
degree of oxygen order derived from the ratio of (3/2 0 0) and (0 3/2
0) oxygen satellite peak intensities equals the twinning ratio of
70$\%$. This shows that the chains are fully oxygen ordered within each
orthorhombic domain.

\begin{figure} [t]
\includegraphics[width=80mm]{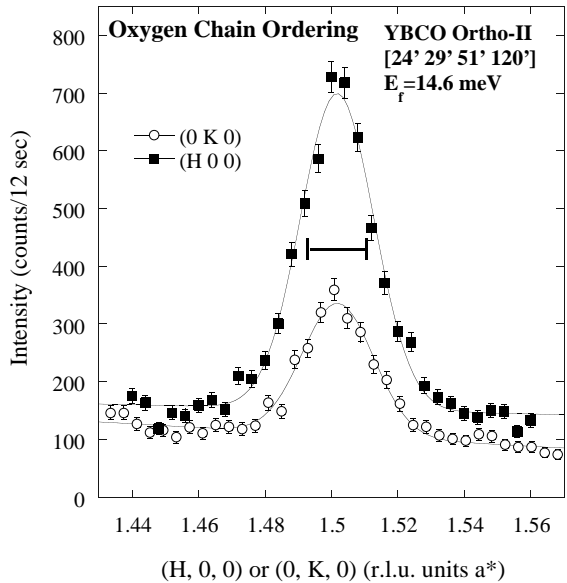}
\caption{Radial scans through the oxygen ordering superlattice peaks at
(3/2 0 0) and (0 3/2 0) (units of a*).  The horizontal bar indicates
the resolution.  A fit to a resolution-convolved lorentzian shows the
oxygen ordering correlation length exceeds 100 \AA.  The larger peak
(filled squares) is from the majority domain and the lesser from the
minority domain.}
\end{figure}

Elastic scattering measurements were carried out at the C5 and E3
neutron spectrometers at the NRU reactor at Chalk River Laboratories
using filtered beams of both 2.37 \AA~and 4 \AA~neutrons.  A focusing
graphite (002) monochromator and a graphite (002) analyzer were used.
Pyrolytic graphite filters (with a total thickness 10 cm) were placed
in the incident and scattered beams to eliminate higher order
reflections.  Before the monochromator, filters of cold sapphire and
beryllium were installed for 2.37 \AA~and 4 \AA~neutron beams
respectively.  For elastic scattering, the collimation was [24$'$ 29$'$
51$'$ 120$'$] horizontally and [80$'$ 240$'$ 214$'$ 429$'$] vertically.
The sample was mounted in a closed-cycle refrigerator on a C-cradle so
that the (H H L) plane was horizontal when the refrigerator was
vertical. The cradle's [001] rotation axis allowed access not only to
(1/2 1/2 L), L=0, 1, 2, where DDW order was sought, but also to (H 0 0)
and (0 K 0) where the detwinning and oxygen order could be measured.

\begin{figure}[t]
\includegraphics[width=80mm]{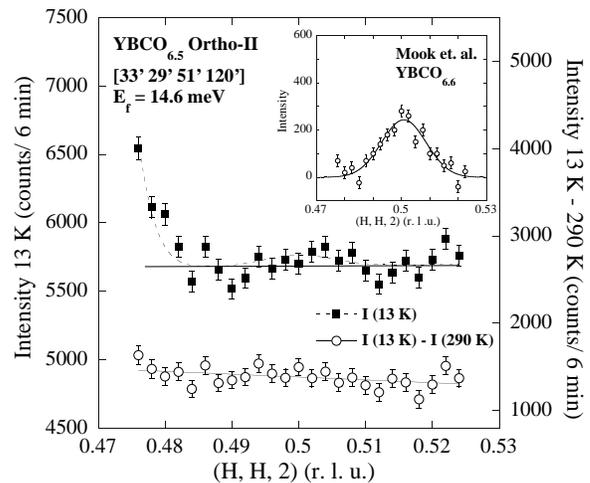}
\caption{Elastic neutron scattering at (1/2 1/2 2) at ($\pi$, $\pi$)
and 13 K shown as filled squares (left hand axis). The rise to the left
is from a green-phase peak.  The data are well described by a flat
background with no sharp magnetic peak (solid line). A gaussian having
the resolution width has been forced to fit the data (broken line) and
yields a sensitivity of $\sim$ 0.003 $\mu_{B}$. The temperature
dependence (open circles and right hand axis) confirms the absence of a
peak. The inset shows the data of Ref.\ \onlinecite{Mook01:64}.}
\end{figure}

For the inelastic neutron scattering measurements at E$_{f}$ = 14.6 meV
and for energy transfers below 10 meV, where the scattering is weaker,
the horizontal collimation was changed to [31$'$ 48$'$ 51$'$ 120$'$],
but was kept the same as the elastic measurements for higher energy
transfers. A pyrolytic graphite filter (5 cm thick) was placed in the
scattered beam.  Inelastic scans were made along the [100], [010], and
[001] directions independently with respect to the zone centers (1/2
1/2 L), (1/2 3/2 L), (1/2 5/2 L).  To do this the rotation axis of the
C-cradle was placed along the [$\overline{1}$10] direction. By slaving
the C-rotation to the L value we could access a general (H K L)
reflection.

\section{Results}
To search for static magnetic order at ($\pi$, $\pi$) we made radial
and transverse elastic scans at the (1/2 1/2 L) positions with L = 0,
1, and 2.  In Fig. 3 we show the scans with 2.37 \AA~neutrons through
(1/2 1/2 2) at 13 K, and the intensity change with temperature in the
normal phase at 290 K, where the peak seen by others has almost
vanished.  The configuration is similar to that of Ref.\
\onlinecite{Mook01:64} and covers the same range of wave vector. We
find, in searches at L=0, 1 and 2, that no static ordering peak exists
in YBa$_{2}$Cu$_{3}$O$_{6.5}$ ortho-II that exceeds about $1.5\%$ of
the 13 K background.  In contrast, the signal for YBCO$_{6.6}$
(results~\cite{Mook01:64} for L=2 are inset in Fig. 3), which lies on a
background of 1850 counts, represents a modulation of $13\%$ above
background, and would have been easily detected.  From the statistical
accuracy of each point alone we would have been sensitive to a moment
at least three times smaller than that reported for
YBCO$_{6.6}$.~\cite{Mook01:64} In a further test (not shown) we again
determined that no peak was present when we used filtered 4
\AA~neutrons in a configuration similar to that of Ref.\
\onlinecite{Sidis01:86}, who reported a much larger moment than that of
Ref.\ \onlinecite{Mook01:64} for a sample of YBCO$_{6.5}$.  We do not
understand the substantial increase of the Q-independent scattering
with decreasing temperature amounting to 300 counts at 70 K and 1300
counts at 290 K.  In the context of the $\sim$300 K transition seen in
other samples, this might signify the growth below room temperature of
a broad peak that extends over much of the Brillouin zone and cannot be
resolved from the background.  It would correspond to highly localized
correlations.

To establish an improved limit on the maximum magnetic moment to which
our experiment is sensitive we put our measurements on an absolute
scale. We calculated our sensitivity in terms of what moment would
produce an intensity corresponding to the average of the error bars in
the Fig. 3 scan through (1/2 1/2 2).  To do this we assumed an
antiferromagnetically ordered array of Cu$^{2+}$ moments pointing along
the \textit{a*} direction (different directions in the a-b plane give a
qualitatively similar estimate).  We compared the average error bar
with the integrated intensity of the H-scan through (3/2 0 0) and the
rock scan through (1 1 2).  These two peaks are ideal for normalization
since they have weak structure factors and should not be subject to
extinction.  These independent methods test the reliability of our
estimate, since the H-scan through (3/2 0 0) requires knowledge of the
instrumental resolution, while the rocking scan through (1 1 2) does
not, as shown by Cowley and Bates.~\cite{Cowley88:21}  We can conclude
that our sensitivity is such that we could have detected any static
moment greater than $\sim$ 0.003 $\mu_{B}$. This is approximately an
order of magnitude less than the moments reported in other
experiments~\cite{Sidis01:86,Mook01:64}. We have also forced a weak
gaussian, with the calculated resolution width, to go through the 13 K
data (Fig. 3) and reached a similar conclusion - moments of the size
observed in these experiments do not occur in well-ordered ortho-II
YBa$_{2}$Cu$_{3}$O$_{6.5}$.

Although we find no static peak we have searched for other signatures
of orbital magnetism.  For example, the DDW theory requires an opening
of a gap in the spin spectrum well above
T$_{c}$.~\cite{Chak01:109,Chak01:204} For this reason it is claimed
that the reported spin gaps $\sim$ 20 meV for YBCO$_{6.6}$ and $\sim$
16 meV for YBCO$_{6.5}$, are a confirmation of the DDW theory. However,
these are superconducting, not normal-phase gaps, since they were
derived by subtracting the spin correlation data above T$_{c}$ from
that below T$_{c}$. In contrast there is no evidence for a spin gap in
disordered YBCO$_{6.5}$ in its normal phase obtained by Fong \textit{et
al.}~\cite{Fong00:61} or Bourges \textit{et al.}~\cite{Bourges95:215}.

Since the presence of a gap would be a key piece of evidence that the
orbital currents of the DDW had broken a discrete symmetry below a
sharp phase transition temperature, we made careful Q-scans in the
low-energy regime to establish whether a spin gap had occurred in the
presence of oxygen order. Previous polarized neutron experiments have
shown that the Q-correlated peak near ($\pi$, $\pi$) is
magnetic,~\cite{Fong97:78} and we have confirmed this from its form
factor and temperature dependence.~\cite{Stock_tobepub} As shown in Fig
4, our experiments provide compelling evidence that a Q-correlated
signal near ($\pi$, $\pi$) persists to the lowest
energies.~\cite{NoteoddL:only} In particular, it exists in the normal
phase for energies well below 20 meV and is still seen at energies as
low as $\sim$ 6 meV in the normal phase. Its spectral weight declines
roughly with E at low energies. We conclude that there is no normal
state spin gap for YBCO$_{6.5}$ ortho-II.

\begin{figure} [t]
\includegraphics[width=80mm]{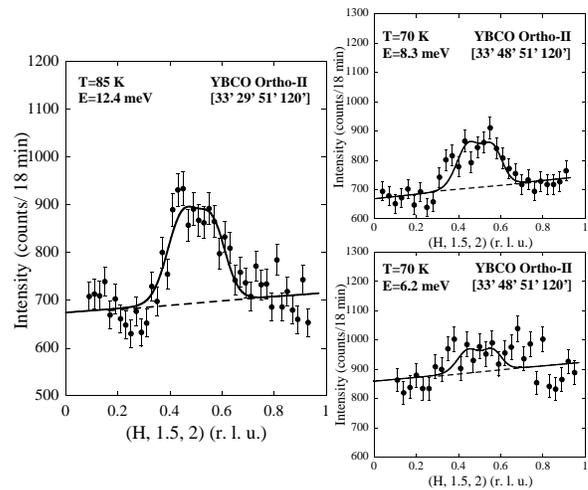}
\caption{Inelastic neutron scattering results in the normal state at
energy transfers of 12.4, 8.3, and 6.2 meV.  Corrections for higher
order contamination have been included. The fits are based on the
over-damped spin-wave model of Chou \textit{et al}. The results show
that Q-correlated spin fluctuations persist to the lowest energies.}

\end{figure}

Our observation that spin fluctuations persist to low energies in
ortho-II ordered YBa$_{2}$Cu$_{3}$O$_{6+x}$ with $x=0.5$ agrees with
the absence of a gap~\cite{Fong00:61,Bourges95:215} in disordered
samples with \textit{x}=0.5 and 0.52.  Relative to these disordered
crystals with similar doping but lower T$_{c}$, the low-energy
susceptibility in the ortho-II crystal is more highly suppressed
relative to the resonance. Our results are also consistent with those
of Chou \textit{et al.}~\cite{Chou91:43} for disordered, twinned
YBCO$_{6.5}$. Our higher resolution and more ordered crystal now
reveals the flat topped correlations in Fig 4, indicative of the
incommensurate structure of the ($\pi$, $\pi$) peak, to be described
elsewhere.~\cite{Stock_tobepub} Otherwise the measured spin response is
similar to that of Chou \textit{et al}. The overdamped spin-wave
model~\cite{Chou91:43} (curves in figure 4) convolved with the
instrumental resolution gives a reasonable description of the data. The
presence of low-energy spin correlations is not consistent a fully
developed spin gap. This makes it unlikely that a new phase has been
entered in which the discrete symmetry of the the Ising-like DDW order
parameter has been broken.~\cite{Chak01:109}

\section{Discussion}

In disordered YBCO the spin fluctuations in YBCO with oxygen
concentrations between 6.4 and 6.5 weaken with hole doping away from
the antiferromagnetic phase and suddenly drop in strength when
\textit{x} reaches 0.5, concomitant with the growth of the orthorhombic
and oxygen chain structure.~\cite{Shirane90:41,Chou91:43}  For ordered
ortho-II with $x=0.5$, we find the spin fluctuations are quite similar
being suppressed but not eliminated.

The sample of YBCO$_{6.6}$ used in Ref.\ \onlinecite{Mook01:64} has
oxygen ordering correlation lengths comparable to ours by comparison of
the peak widths.~\cite{Mook:priv}  However the extra oxygen must find a
way into the lattice and interrupt the perfect 2\textit{a} by
1\textit{b} cell of the ortho-II structure.  The sample of Ref.\
\onlinecite{Sidis01:86} has shorter oxygen correlation lengths of 20
\AA~in the \textit{a-b} plane and 12 \AA~along the \textit{c}
direction. These are shorter than in our sample, which exhibits nearly
resolution limited superlattice peaks indicating oxygen order over more
than $\sim100$ \AA~within the CuO$_{2}$ plane and 50 \AA~normal to the
planes. This might suggest that disorder causing the proximity of
adjacent short segments of chains filled with oxygen could lead to
magnetic order.

The idea that disorder may induce long-range antiferromagnetic order is
not new.  Recent examples include Mg doped CuGeO$_{3}$ which exhibits
long-range antiferromagnetic order for finite Mg
concentration.~\cite{Nakao99:68}  Also Hodges \textit{et
al.}~\cite{Hodges02:218} added only $1.3\%$ of Co to optimally doped
YBCO and found long-range antiferromagnetic order below a large
temperature, T = 320 K, quite similar to the onset of elastic peaks in
refs.\ \onlinecite{Sidis01:86} and\ \onlinecite{Mook01:64}.  They
suggest this impurity effect is analogous to the order seen by Sidis
\textit{et al.}~\cite{Sidis01:86} in underdoped YBCO.  The latter
example is particularly interesting as Co is known to enter into the
copper chains and pull into the structure 0.5 oxygen atoms per Co atom,
therefore introducing disorder into the chains~\cite{Hodges02:218}.

A number of theories describing the effect of disorder and impurities
have been formulated.~\cite{Tanaka01:071,Vojta00:61} In the
Ginsburg-Landau analysis by Kohno \textit{et
al.}\cite{Kohno990209:Feb99} disorder in a region can weaken the
superconducting order and allow antiferromagnetic correlations to grow
locally. With increasing disorder or impurity concentration the
superconducting state will be transformed first into a state with
locally nucleated antiferromagnetic moments and later into a state of
long-range antiferromagnetic order. We believe our results represent
the former state of weak disorder where there is coexistence of
moderately suppressed superconductivity with only short-range magnetic
order. Systems where a phase has been found below a well-defined
transition temperature and where a sharp DDW peak occurs, might then
lie in the latter region; here the nature of the disorder creates
long-range magnetic order while more strongly suppressing
superconductivity. We are therefore closer to the clean system referred
to by Kohno \textit{et al}.

It is possible that ortho-II YBCO$_{6.5}$ lies in the phase diagram at
an oxygen doping where the low-energy spin susceptibility, being the
sub-critical fluctuations of the antiferromagnetic quantum critical
point, is strongly suppressed by the growth of orbital current
fluctuations.  However, there is no occurrence of static DDW order at a
sharp temperature accompanied by a spin gap. Our results for
\textit{x}=0.5 are consistent with the work of Shirane \textit{et
al.}~\cite{Shirane90:41}, Chou \textit{et al.}~\cite{Chou91:43} and
Regnault \textit{et al.}~\cite{Regnault94:238} who have shown that as
\textit{x} increases the low-energy weight of $\chi''(\omega)$ is
suppressed but not eliminated. Taken in conjunction with the evidence
at other hole dopings, we believe there is no sharp transition to a
gapped state, but rather a crossover with increasing \textit{x} to a
state where low-energy spin fluctuations carry little weight but could
coexist with the fluctuations of a different symmetry. This idea
receives support in the recent work of Lee \textit{et
al.}~\cite{Lee02:052} who propose that the pseudogap state is not
characterized by a phase transition but by a cross-over to a state with
strong fluctuations between orbital currents and d-wave
superconductivity.

\section{Conclusion}

Static antiferromagnetic long-range correlations that would produce a
peak at ($\pi$, $\pi$) are absent in YBCO6.5 in its ortho-II structure
at a level that represents our magnetic moment sensitivity of 0.003
$\mu_{B}$. This is more than an order of magnitude lower than that of
the elastic peaks reported in underdoped cuprate superconductors with
somewhat different oxygen doping and/or order.  We cannot exclude the
possibility that a broad peak may exist with extremely short-range DDW
correlations.  The generic phase diagram~\cite{Chak01:204}, may allow
YBCO at larger doping such as  O$_{6.6}$ to lie in the DDW phase while
O$_{6.5}$ does not.

One possible way to reconcile the differences between experiments is
that structural disorder (of the oxygens, for example) induces a phase
transition to a long-ranged ordered state.  The absence of sharp static
peaks contrasts with theoretical predictions for static orbital
currents or d-density-wave states. We cannot exclude an orbital current
state that does not break the symmetry of the Cu$_{2}$O
planes.\cite{Varma97:55} Spin fluctuations do exist near ($\pi$, $\pi$)
but they are dynamic and are found to extend in the normal phase to low
energies ($\sim$ 6 meV). Instead of the prediction~\cite{Chak01:63}
that the spin-fluctuation spectrum ``remains fully gapped and has no
low-energy structure of any kind" we find that the low-energy response
is suppressed but not eliminated. Our results do not exclude the
possibility that the superconductivity in underdoped YBCO coexists with
dynamic orbital currents, an idea that has received recent theoretical
support.~\cite{Lee02:052}

\section*{ACKNOWLEDGMENTS}

We are grateful to H.A. Mook who kindly supplied his values of
correlation lengths prior to publication and made helpful comments on
the manuscript, and  also to P. Dai and S. Chakravarty. Mark Bentall,
Oxford, helped with part of the experiments. We had valuable
discussions with R. J. Birgeneau, R.A. Cowley, P. Clegg, J. S. Gardner
and Z. Yamani. We would especially like to thank S. Wakimoto for useful
discussions and for a critical reading of the manuscript. We are
grateful to L. McEwan, R. L. Donaberger, J. H. Fox, and M. M. Potter
for technical assistance. The work at the University of Toronto was
supported by NSERC.




\begin{thebibliography}{}
\bibitem{NoteoddL:only} This result holds for the odd channel, L$\sim$2,
where the spin correlation between the bilayers are antiferromagnetic.
\end{thebibliography}
\end{document}